\documentclass[10pt]{amsart}
\usepackage{amssymb}
\usepackage{enumerate}
\usepackage[dvips]{graphicx}
\DeclareGraphicsExtensions{.eps,.art,.ART,.ps}
\newcommand{\bbR}{\mathbb{R}}      

\newcommand{\tr}{\operatorname{tr}}
\newcommand{\grad}{\operatorname{grad}}
\newcommand{\curl}{\operatorname{curl}}
\newcommand{\dive}{\operatorname{div}}
\theoremstyle{definition}

\theoremstyle{remark}

\begin{document}
\begin{abstract}
We write the Mathisson-Papapetrou equations of motion for a spinning particle in a stationary spacetime using the quasi-Maxwell formalism and give an interpretation of the coupling between spin and curvature. The formalism is then used to compute equilibrium positions for spinning particles in the NUT spacetime.
\end{abstract}
%
%
\title{Quasi-Maxwell interpretation of the spin-curvature coupling}
\author{Jos\'{e} Nat\'{a}rio}
\address{Department of Mathematics, Instituto Superior T\'{e}cnico, Portugal}
\thanks{This work was partially supported by FCT/POCTI/FEDER}
\maketitle
%
%
%
\section*{Introduction}
The purpose of this paper is to use the quasi-Maxwell form of Einstein's field equations for a stationary spacetime to give an interpretation of the force that acts on a spinning particle according to the Mathisson-Papapetrou equations.

The paper is divided into five sections. In the first section, we briefly review the quasi-Maxwell formalism for stationary spacetimes. The Mathisson-Papapetrou equations of motion for a spinning particle, as well as the various spin supplementary conditions, are discussed in the second section. The third section recalls the main formulae for a magnetic dipole on a magnetic field, which are compared with their gravitational counterparts in the fourth section. It turns out that there is a remarkable analogy between the two sets of formulae, shedding light into the nature of the spin-curvature coupling. In the fifth section, as an example, we use the quasi-Maxwell formalism to compute equilibrium positions for spinning particles in the NUT spacetime.

We use the notation and sign conventions in \cite{W84}, where latin indices represent abstract indices, except for the latin indices $i,j,\ldots$ , which will be {\em numerical} indices referring to the space manifold (i.e. $i,j,\ldots = 1,2,3$). As is usual, we will not worry about the vertical position of space indices on orthonormal frames.
%
%
%
\section{Quasi-Maxwell formalism for stationary spacetimes}
We start by briefly reviewing the quasi-Maxwell formalism for stationary spacetimes. For more details, see \cite{O02, CN05, Embacher84, LL97, LN98, NZ97, BS00}.

Let $(M,g)$ be a chronological stationary spacetime with a complete timelike Killing vector field $T$. Then the action of $\bbR$ on $M$ by the flow of $T$ makes it a trivial principal $\bbR$-bundle \cite{An00, BS00}, and the quotient space $M / \bbR$ can be identified with a $3$-dimensional submanifold $\Sigma \subset M$. One can think of this quotient as the set of all stationary observers. If $\pi:M \to \Sigma$ is the quotient map, $p \in \Sigma$ and $v \in T_p \Sigma$, we define $v^\dagger$ to be the vector field along the integral line of $T$ through $p$ which is orthogonal to $T$ and such that $\pi_* v^\dagger = v$. We can then define the Riemannian metric $h$ on $\Sigma$ through
\[
h(v,v) = g(v^\dagger, v^\dagger).
\]
This metric has the physical interpretation of yielding the distance measured between nearby stationary observers through Einstein's light signaling procedure \cite{LL97}. The {\bf space manifold} is the Riemannian manifold $(\Sigma, h)$.

There exists a function $\phi: \Sigma \to \bbR$, called the {\bf gravitational potential}, such that
\[
g(T,T)=-e^{2 (\phi \, \circ \, \pi)}.
\]
The {\bf gravitational field} is then the vector field ${\bf G}$ defined on $\Sigma$ by
\[
{\bf G}=-\grad \phi.
\]
It is easy to show that $-{\bf G}^\dagger$ is simply the acceleration of the stationary observers.

For a particular choice of the submanifold $\Sigma \subset M$, one can define the one-form $A$ through
\[
A(v)=-e^{-2\phi}g(T,v)
\]
for all $v \in T \Sigma \subset TM$. It turns out that although $A$ depends on the choice of $\Sigma$, the $2$-form
\[
H = - e^{\phi} dA
\]
does not, and hence is well defined on the quotient. Assuming that $\Sigma$ is orientable, we define the {\bf gravitomagnetic field} to be the vector field 
\[
{\bf H} = (^*H)^\sharp,
\]
where $^*$ is the Hodge star and $^\sharp$ is the usual map between vectors and covectors determined by $h$. It is not hard to show that $\frac12 {\bf H}^\dagger$ is simply the vorticity of the stationary observers.

The names above are justified by the fact that the projection of any timelike geodesic on the space manifold, parameterized by the proper time $\tau$, satisfies the equation
\[
\frac{D {\bf u}}{d\tau} = \gamma^2 {\bf G} + \gamma {\bf u} \times {\bf H},
\]
where ${\bf u}$ is the tangent vector and $\gamma = \left( 1 + {\bf u}^2 \right)^\frac12$ is the energy per unit mass of the corresponding free-falling particle as measured by stationary observers. As a consequence of this equation, the particle's total energy per unit mass
\[
E = e^\phi \gamma
\]
is conserved.

If $\{E_1, E_2, E_3\}$ is a right-handed orthonormal local frame on the space manifold, then $\{E_0, E_1^\dagger, E_2^\dagger, E_3^\dagger \}$ is an orthonormal local frame on $M$, where $E_0 = e^{-\phi} T$. The nonvanishing Christoffel symbols associated to this frame are \cite{O02}
\begin{align*}
& ^{(4)}\Gamma^i_{00} = ^{(4)}\Gamma^0_{0i} = - G_i; \\
& ^{(4)}\Gamma^0_{ij} = - ^{(4)}\Gamma^i_{0j} = - ^{(4)}\Gamma^i_{j0} = \frac12 H_{ij}; \\
& ^{(4)}\Gamma^i_{jk} = \Gamma^i_{jk},
\end{align*}
where $\Gamma^i_{jk}$ are the Christoffel symbols associated to $\{E_1, E_2, E_3\}$. The components of the Riemann curvature tensor are therefore given by
\begin{align*}
& ^{(4)}R_{0i0j} = - \nabla_j G_i + G_i G_j - \frac14 H_{ik} H_{kj}; \\
& ^{(4)}R_{0ijk} = \frac12 \left( \nabla_j H_{ik} - \nabla_k H_{ij} - 2 G_i H_{jk}\right); \\
& ^{(4)}R_{ijkl} = R_{ijkl} + \frac14 \left( 2 H_{ij} H_{kl} + H_{ik}H_{jl} - H_{il}H_{jk} \right),
\end{align*}
and the components of the Ricci curvature tensor are
\begin{align*}
& ^{(4)}R_{00} = - \dive {\bf G} + {\bf G}^2 + \frac12 {\bf H}^2; \\
& ^{(4)}R_{0i} = \frac12 \left(\curl {\bf H}\right)_i - \left({\bf G} \times {\bf H}\right)_i; \\
& ^{(4)}R_{ij} = R_{ij} + \nabla_i G_j - G_i G_j - \frac12 H_i H_j + \frac12 {\bf H}^2 h_{ij}.
\end{align*}
In particular, one can write Einstein's equations in the quasi-Maxwell form
\begin{align*}
& \dive {\bf G} = {\bf G}^2 + \frac12 {\bf H}^2 - 4\pi\left(\rho + \tr \sigma \right); \\
& \curl {\bf H} = 2{\bf G} \times {\bf H} - 16 \pi {\bf j}; \\
& Ric + \nabla {\bf G}^\sharp = {\bf G}^\sharp \otimes {\bf G}^\sharp + \frac12 {\bf H}^\sharp \otimes {\bf H}^\sharp - \frac12 {\bf H}^2 h + 8 \pi \sigma + 4\pi\left(\rho - \tr \sigma \right) h,
\end{align*}
where $Ric$ is the Ricci curvature of the space manifold and $\rho = T^{00}$, ${\bf j} = T^{0i} E_i$ and $\sigma=T_{ij}E_i^\sharp \otimes E_j^\sharp$ are the energy density, the energy current and the spatial stress tensor measured by the stationary observers. These equations imply that the scalar curvature of the space manifold is
\[
R = 16 \pi \rho - \frac32 {\bf H}^2.
\]
%
%
%
\section{Spinning particle in General Relativity}
The Mathisson-Papapetrou equations \cite{Mathisson37, P51} for the motion of a spinning particle, obtained by neglecting all multipoles higher than mass monopole and spin dipole \cite{D70, ADS03}, are
\begin{align*}
& \nabla_U P^a + \frac12 R^a_{\,\,\,\, bcd} U^b S^{cd} = 0 \\ 
& \nabla_U S - P \wedge U = 0
\end{align*}
where $U$ is the unit tangent vector to the particle's history, $P$ is its energy-momentum vector (not necessarily parallel to $U$) and $S$ is a rank-$2$ antisymmetric tensor representing the particle's angular momentum. This system of equations is underdetermined, and must be closed by means of a {\bf spin supplementary condition}. The usual choice, justified in part by the uniqueness results in \cite{B67, S79a, S79b}, is the Tulczyjew-Dixon condition \cite{Tulczyjew59, D70}
\[
S^{ab} P_b = 0.
\]
Less popular choices include the Mathisson-Pirani condition \cite{Mathisson37, Pirani56}
\[
S^{ab} U_a = 0
\]
(for recent reviews of these matters see for instance \cite{dFC95, S99}; recent work on solutions of these equations in Schwarzschild and Kerr spacetimes can be found in \cite{SM96, S99, H03, BdFG04a, BdFG04b, Singh05, MS06}).

Dixon's supplementary condition implies that the angular momentum tensor is of the form
\[
S^{ab} = \frac1m \epsilon^{abcd} P_c s_d,
\]
where $\epsilon$ is the volume element and $m = \sqrt{-g(P,P)}$ is the particle's {\bf dynamical mass} (which can be seen to be constant). The {\bf spin vector}
\[
s^a = - \frac1{2m} \epsilon^a_{\,\,\,\, bcd} P^b S^{cd}
\]
(which is the relativistic analogue of the angular momentum vector) is orthogonal to $P$ (hence spacelike), and satisfies
\[
\nabla_U s = \frac1{m^2} g\left( \nabla_U P, s \right) P.
\]

It is possible to show \cite{Kunzle72, TdFC76, S99} that
\begin{equation} \label{relation}
U^a = \frac{g(P, U)}{m^2} \left( P^a + \frac{1}{\Delta} S^{ab}R_{bcde}P^cS^{de} \right),
\end{equation}
where
\[
\Delta = 2m^2 + \frac12 R_{abcd} S^{ab} S^{cd}.
\]
Therefore $U$ and $\frac1m P$ differ by a Lorentz transformation of order
\[
\frac{|s|^2}{m^2 r^2},
\] 
where $r$ is the local radius of curvature. Assuming that the particle is not rotating too fast, we can take $P$ and $U$ to be parallel, $P = mU$, in which case Dixon's and Mathisson's spin supplementary conditions coincide. The (approximate) Mathisson-Papapetrou equations are then simply written as
\begin{align}
& m\nabla_U U^a = - \frac12 R^a_{\,\,\,\, bcd} U^b \epsilon^{cdef} U_e s_f; \label{approximate1} \\ 
& \nabla_U s = g\left( \nabla_U U, s \right) U. \label{approximate2}
\end{align}
These equations are consistent with the spin supplementary condition $g(U,s)=0$, as $s$ is Fermi-Walker transported along the motion.

The first equation indicates that a net force arises from the particle's spin coupling to the spacetime curvature. The expression for this force, however, is not particularly enlightening. We will see that within the quasi-Maxwell formalism its expression is actually quite natural.
%
%
%
\section{Magnetic dipole on a magnetic field}
As is well known (see for instance \cite{J98}), a magnetic dipole with moment ${\bf m}$ placed in a magnetic field ${\bf B}$ will experience a torque
\[
{\bf N} = {\bf m} \times {\bf B},
\]
which can be thought of as arising from the potential energy
\[
U = - {\bf m} \cdot {\bf B}.
\]
Such dipole will experience a net force
\[
{\bf F} =  \nabla{\bf B} \cdot {\bf m} - \left(\dive {\bf B}\right) {\bf m},
\]
where the last term is not usually written because of Maxwell's equation $\dive {\bf B}=0$.
%
%
%
\section{Spinning particle in Quasi-Maxwell formalism}
We will consider the special case when the particle is at rest with respect to the stationary observers, i.e.
\[
U = E_0.
\]
The spin supplementary condition then yields
\[
s^0 = 0.
\]
Using the formulae for the Christoffel symbols, it is easily seen that the equation for the Fermi-Walker transport of $s$ can be written as
\begin{align*}
\frac{ds^i}{d\tau} + ^{(4)}\Gamma^i_{0j} s^j = 0 \Leftrightarrow  \frac{ds^i}{d\tau} - \frac12 H_{ij} s^j = 0,
\end{align*}
i.e.
\begin{equation} \label{prec}
\frac{D{\bf s}}{d\tau} = - \frac12 {\bf H} \times {\bf s}.
\end{equation}
This formula shows that the particle's angular momentum changes under a gravitomagnetic field exactly like a magnetic dipole moment under a magnetic field (except for a factor $\frac12$).

Using the expressions for the components of the Riemann tensor, one can compute the force due to coupling between spin and curvature to have components
\[
F^i = - \frac12 \, ^{(4)}R^i_{\,\,\,\, 0jk} \epsilon^{jk0l} (-1) s_l = \frac14 \left(\nabla_j H_{ik} - \nabla_k H_{ij} - 2G_i H_{jk}\right) \varepsilon_{jkl} s^l
\]
where $\varepsilon$ is the volume element of the space manifold\footnote{We have $\epsilon^{0ijk}=-\varepsilon_{ijk}$ on any orthonormal frame.}. Therefore
\begin{align*}
F^i & = \frac12 \left(\nabla_j H_{ik} - G_i H_{jk}\right) \varepsilon_{jkl} s^l = \frac12 \left(\nabla_j ( \varepsilon_{ikm} H^m ) - G_i \varepsilon_{jkm} H^m \right) \varepsilon_{jkl} s^l \\
& = \frac12 \left( \delta_{ij} \delta_{ml} - \delta_{il} \delta_{mj} \right) (\nabla_j H^m) s^l - \frac12 \left( \delta_{jj} \delta_{ml} - \delta_{jl} \delta_{mj} \right) G_i H^m s^l \\
& = \frac12 (\nabla_i H^l) s^l - \frac12 (\nabla_j H^j) s^i - \frac12 \left( 3\delta_{ml} - \delta_{ml} \right) G_i H^m s^l \\
& = \frac12 (\nabla_i H^j) s^j - \frac12 (\nabla_j H^j) s^i - (H^j s^j) G_i,
\end{align*}
that is,
 \begin{equation} \label{F}
{\bf F} = \frac12 \nabla{\bf H} \cdot {\bf s} - \frac12 \left(\dive {\bf H}\right) {\bf s} - \left({\bf s} \cdot {\bf H}\right) {\bf G}.
\end{equation}

Therefore the spin-curvature coupling can be interpreted as consisting of a gravitomagnetic part, which mimics the magnetic force on a magnetic dipole (apart from the factor $\frac12$), plus the weight of the dipole energy.

This similarity between the spin-curvature coupling and the force on a magnetic dipole was noticed long ago in the context fo the linearized theory by Wald \cite{W72}. Equation~(\ref{F}) extends this analogy to an arbitrarily strong (albeit stationary) field. Notice that only the first term in (\ref{F}) appears in \cite{W72}, as the other two are not linear in the fields (it is easily seen that $\dive {\bf H} = -{\bf G} \cdot {\bf H}$).
%
%
%
\section{Spinning particle in NUT spacetime}
We would like to use equation~(\ref{F}) to compute equilibrium positions for spinning particles in stationary spacetimes. Unfortunately, this is a situation where the approximate equations (\ref{approximate1}) and (\ref{approximate2}) are expected to break down. Indeed, since the components of the Riemann tensor are quadratic in the components of the fields ${\bf G}$ and ${\bf H}$, we expect these to be of order $\frac1r$, where $r$ is the local radius of curvature. Therefore the force on the spinning particle is of order
\[
\frac{|s|}{r^2}. 
\]
For the particle to remain at equilibrium, this force must balance the gravitational force $m{\bf G}$, which is of order $\frac{m}{r}$. Therefore, we expect
\[
\frac{|s|}{r^2} \sim \frac{m}{r} \Leftrightarrow \frac{|s|}{mr} \sim 1
\]
at any equilibrium position, whereas the approximate equations assume that this quantity is small.

Equation (\ref{relation}), however, implies that the relation $U = \frac1m P$ is exact if ${\bf s} \times {\bf F} = {\bf 0}$, where ${\bf F}$ is the force on the spinning particle given in (\ref{F}); for an equilibrium position, this is equivalent to requiring that ${\bf s} \times {\bf G} = {\bf 0}$.\footnote{It is interesting to note that the linear momentum of the field produced by a magnetic dipole ${\bf m}$ on an electric field ${\bf E}$ has the approximate expression ${\bf P}_{\text{field}} = {\bf E} \times {\bf m}$ \cite{J98}.} Therefore the equilibrium positions computed using equation (\ref{F}) will be exact for the Tulczyjew-Dixon supplementary condition if the angular momentum is aligned with the gravitational field.

For completeness, let us remark that the Mathisson-Pirani spin supplementary condition leads to the motion equation
\[
m \nabla_U U^a = - \frac12 R^a_{\,\,\,\, bcd} U^b S^{cd} + S^{ab} \nabla_U \nabla_U U_b,
\]
where the constant $m$ is now defined as $m=-g(P,U)$ \cite{Mathisson37, Pirani56, MTW73, Stephani04}. For an equilibrium position, this will reduce to the approximate equation (\ref{approximate1}) when ${\bf s} \times ({\bf G} \times {\bf H}) = {\bf 0}$. Therefore the equilibrium positions computed using equation (\ref{F}) will be exact for the Mathisson-Pirani supplementary condition if the angular momentum is aligned with the ``Poynting vector'' ${\bf G}\times{\bf H}$. Thus for instance the equilibrium positions along the axis of  the Kerr-de Sitter solution, computed in \cite{SK06} using the Mathisson-Pirani supplementary condition, coincide with the equilibrium positions computed using Tulczyjew-Dixon supplementary condition, and can both be obtained from (\ref{F}).

As an example, we now compute equilibrium positions for spinning particles in NUT spacetime. Recall that the NUT spacetime is a stationary solution of the vacuum Einstein field equations describing a gravitomagnetic monopole, given in local coordinates by
\[
g = - e^{2\phi} (dt + 2l \cos \theta d\varphi)^2  + e^{-2\phi} dr^2 + (r^2 + l^2) (d \theta^2 + \sin^2 \theta d \varphi^2),
\]
where
\[
e^{2\phi} = 1 - 2 \frac{Mr + l^2}{r^2 + l^2}
\]
and $M,l$ are two parameters representing the total mass and (half) the gravitomagnetic charge \cite{NUT63, LN98, SKMHH03}. The metric of the space manifold determined by the stationary observers is
\[
h = e^{-2\phi} dr^2 + (r^2 + l^2) (d \theta^2 + \sin^2 \theta d \varphi^2),
\]
and the gravitomagnetic potential $1$-form is
\[
A = 2l \cos \theta d\varphi.
\]
Thus the gravitomagnetic field corresponds to the $2$-form
\[
H = - e^\phi dA = 2l e^\phi \sin \theta d\theta \wedge d \varphi,
\]
and hence to the $1$-form
\[
{\bf H}^\sharp = ^*H = \frac{2l}{r^2 + l^2} dr,
\]
implying that ${\bf H}$ is radial. If we choose the angular momentum ${\bf s}$ to be also radial, then the equilibrium positions will be exact for the Tulczyjew-Dixon supplementary condition. Moreover, since the $1$-form associated to the gravitational field is
\[
{\bf G}^\sharp = - d\phi = - \phi' dr,
\]
where the prime denotes differentiation with respect to $r$, we have ${\bf G} \times {\bf H} = 0$, and the equilibrium positions will also be exact for the Mathisson-Pirani supplementary condition.

By equation (\ref{prec}), the angular momentum ${\bf s}$ will be constant. We now compute the $1$-forms associated to each term on the expression (\ref{F}) for the force on a spinning particle (here we must keep track of the vertical position of the indices, as we are working with a non-orthonormal coordinate basis). To begin with, we have
\[
(\nabla{\bf H} \cdot {\bf s})^\sharp = \nabla_r H_r s^r dr = (H_r' - \Gamma^r_{rr} H_r) s^r dr = (H_r' + \phi' H_r) s^r dr
\]
(the relevant Christoffel symbols can be readily computed from the geodesic equations of the space manifold). Moreover,
\[
(\dive {\bf H}) {\bf s}^\sharp = - ({\bf G} \cdot {\bf H}) s_r dr = - G^r H_r s_r dr = - G_r H_r s^r dr = \phi' H_r s^r dr.
\]
Finally,
\[
({\bf s} \cdot {\bf H}) {\bf G}^\sharp = - s^r H_r \phi' dr.
\]
Collecting these terms together yields
\[
{\bf F}^\sharp = \frac12 (H_r' + 2\phi' H_r) s^r dr.
\]
The equation for the equilibrium positions is then
\[
m {\bf G}^\sharp + {\bf F}^\sharp = 0 \Leftrightarrow - 2m\phi' + (H_r' + 2\phi' H_r) s^r = 0 \Leftrightarrow \frac{s^r}{m} = \frac{(e^{2\phi})'}{(H_r e^{2\phi})'}.
\]
For $r \gg \sqrt{M^2 + l^2}$, we have
\[
e^{2\phi} \simeq 1 - \frac{2M}{r}, \quad \quad \quad H_r e^{2\phi} \simeq \frac{2l}{r^2},
\]
and hence the equilibrium condition reduces to
\[
\frac{s^r}{m} \simeq - \frac{Mr}{2l}.
\]
In particular, the angular momentum must point towards the center if $l>0$. This could easily be computed from the linearized formulae in \cite{W72}. In the strong field region, however, things are more complicated. For instance, near the horizon $r_H = M + \sqrt{M^2 + l^2}$ one has
\[
e^{2 \phi} \simeq 0
\]
and hence the equilibrium condition reduces to
\[
\frac{s^r}{m} \simeq \frac{(e^{2\phi})'}{H_r(e^{2\phi})'} = \frac{1}{H_r} = \frac{r^2 + l^2}{2l}.
\]
In particular, the angular momentum must point {\em away} from the center in this region.
%
%
%


\begin{thebibliography}{BdFG04b}

\bibitem[ADS03]{ADS03}
J.~Anandan, N.~Dadhich, and P.~Singh, \emph{Action principle formulation for
  the motion of extended bodies in general relativity}, Phys. Rev. D
  \textbf{68} (2003), 124014.

\bibitem[And00]{An00}
M.~Anderson, \emph{On stationary vacuum solutions to the Einstein equations},
  Annales Henri Poincare \textbf{1} (2000), 977--994.

\bibitem[BdFG04a]{BdFG04b}
D.~Bini, F.~de~Felice, and A.~Geralico, \emph{Spinning test particles and clock
  effect in Kerr spacetime}, Class. Quant. Grav. \textbf{21} (2004),
  5441--5456.

\bibitem[BdFG04b]{BdFG04a}
\bysame, \emph{Spinning test particles and clock effect in Schwarzschild
  spacetime}, Class. Quant. Grav. \textbf{21} (2004), 5427--5440.

\bibitem[Bei67]{B67}
Beiglb\"ock, \emph{The center-of-mass in Einstein's theory of gravitation},
  Commun. Math. Phys. \textbf{5} (1967), 106--130.

\bibitem[BS00]{BS00}
R.~Beig and B.~Schmidt, \emph{Time-independent gravitational fields}, Lect.
  Notes Phys. \textbf{540} (2000), 325--372.

\bibitem[CN05]{CN05}
J.~Costa and J.~Nat\'ario, \emph{Homogeneous cosmologies from the quasi-Maxwell
  formalism}, J. Math. Phys \textbf{46} (2005), 082501.

\bibitem[dFC95]{dFC95}
F.~de~Felice and J.~Clarke, \emph{Relativity on curved manifolds}, Cambridge
  University Press, 1995.

\bibitem[Dix70]{D70}
W.~Dixon, \emph{Dynamics of extended bodies in general relativity i. Momentum
  and angular momentum}, Proc. R. Soc. London A \textbf{314} (1970), 499.

\bibitem[Emb84]{Embacher84}
F.~Embacher, \emph{The analog of electric and magnetic fields in stationary
  gravitational systems}, Found. Phys. \textbf{14} (1984), 721--738.

\bibitem[Har03]{H03}
N.~Hartl, \emph{Dynamics of spinning test particles in Kerr spacetime}, Phys.
  Rev. D \textbf{67} (2003), 024005.

\bibitem[Jac98]{J98}
J.~Jackson, \emph{Classical electrodynamics}, Wiley, 1998.

\bibitem[K72]{Kunzle72}
H.~K\"unzle, \emph{Canonical dynamics of spinning particles in gravitational
  and electromagnetic fields}, J. Math. Phys. \textbf{13} (1972), 739--744.

\bibitem[LBNZ98]{LN98}
D.~Lynden-Bell and M.~Nouri-Zonoz, \emph{Classical monopoles: Newton,
  NUT-space, gravomagnetic lensing and atomic spectra}, Rev. Mod. Phys
  \textbf{70} (1998), 427--446.

\bibitem[LL97]{LL97}
L.~Landau and E.~Lifshitz, \emph{The classical theory of fields},
  Butterworth-Heinemann, 1997.

\bibitem[Mat37]{Mathisson37}
M.~Mathisson, \emph{Neue mechanik materieller systeme}, Acta Phys. Polon.
  \textbf{6} (1937), 163--200.

\bibitem[MS06]{MS06}
B.~Mashhoon and D.~Singh, \emph{Dynamics of extended spinning masses in a
  gravitational field}, Phys. Rev. D \textbf{74} (2006), 124006.

\bibitem[MTW73]{MTW73}
C.~Misner, K.~Thorne, and J.~A. Wheeler, \emph{Gravitation}, Freeman, 1973.

\bibitem[NTU63]{NUT63}
E.~Newman, L.~Tamburino, and T.~Unti, \emph{Empty-space generalization of the
  Schwarzschild metric}, J. Math. Phys. \textbf{4} (1963), 915--923.

\bibitem[NZ97]{NZ97}
M.~Nouri-Zonoz, \emph{Cylindrical analogue of NUT space: spacetime of a line
  gravomagnetic monopole}, Class. Quant. Grav. \textbf{14} (1997), 3123--3129.

\bibitem[Oli02]{O02}
W.~Oliva, \emph{Geometric mechanics}, Springer, 2002.

\bibitem[Pap51]{P51}
A.~Papapetrou, \emph{Spinning test particles in general relativity i}, Proc. R.
  Soc. London A \textbf{209} (1951), 248.

\bibitem[Pir56]{Pirani56}
F.~Pirani, \emph{On the physical significance of the Riemann tensor}, Acta
  Phys. Polon. \textbf{15} (1956), 389--405.

\bibitem[Sch79a]{S79a}
R.~Schattner, \emph{The center of mass in general relativity}, Gen. Rel. Grav.
  \textbf{10} (1979), no.~5, 377--393.

\bibitem[Sch79b]{S79b}
\bysame, \emph{The uniqueness of the center of mass in general relativity},
  Gen. Rel. Grav. \textbf{10} (1979), no.~5, 395--399.

\bibitem[Sem99]{S99}
Semer\'ak, \emph{Spinning test particles in a Kerr field - i}, Mont. Not.
  Astron. Soc. \textbf{308} (1999), 863--875.

\bibitem[Sin05]{Singh05}
D.~Singh, \emph{The dynamics of a classical spinning particle in Vaidya
  space-time}, Phys. Rev. D \textbf{72} (2005), 084033.

\bibitem[SK06]{SK06}
Z.~Stuchlik and J.~Kovar, \emph{Equilibrium conditions of spinning test
  particles in Kerr-de Sitter spacetimes}, Class. Quant. Grav. \textbf{23}
  (2006), 3935--3949.

\bibitem[SKM03]{SKMHH03}
H.~Stephani, D.~Kramer, M.~MacCallum, C.~Hoensalaers, and E.~Herlt, \emph{Exact
  solutions of Einstein's field equations}, Cambridge University Press, 2003.

\bibitem[SM96]{SM96}
S.~Suzuki and K.~Maeda, \emph{Chaos in Schwarzschild space-time: The motion of
  a spinning particle}, Phys. Rev. D \textbf{55} (1996), 4848--4859.

\bibitem[Ste04]{Stephani04}
H.~Stephani, \emph{Relativity}, Cambridge University Press, 2004.

\bibitem[TdFC76]{TdFC76}
K.~Tod, F.~de~Felice, and M.~Calvani, \emph{Spinning test particles in the
  field of a black hole}, Nuovo Cim. B \textbf{34} (1976), 365.

\bibitem[Tul59]{Tulczyjew59}
W.~Tulczyjew, \emph{Motion of multipole particles in general relativity
  theory}, Acta Phys. Polon. \textbf{18} (1959), 393--409.

\bibitem[Wal72]{W72}
R.~Wald, \emph{Gravitational spin interaction}, Phys. Rev. D \textbf{6} (1972),
  no.~2, 406--413.

\bibitem[Wal84]{W84}
\bysame, \emph{General relativity}, University of Chicago Press, 1984.

\end{thebibliography}
\end{document}